\documentstyle[aps,prl,multicol]{revtex}

\input epsf.sty

\begin{document}
\draft

\title
{
	Modeling Grain Boundaries using a 
	Phase Field Technique
}
\author{ Ryo Kobayashi}
\address{Research Institute for Electronic Science\\
        Hokkaido University\\ Sapporo 060, Japan}
\author{James A. Warren}
\address{Metallurgy Division and Center for\\Theoretical and
Computational Materials Science\\
National Institute of Standards and Technology\\
Gaithersburg, MD 20899}
\author{W. Craig Carter}
\address{Department of Materials Science and Engineering\\
Massachusetts Institute of Technology\\
Cambridge, MA, 02139-4307}
\date{\today}

\maketitle

\date{\today}

\begin{abstract}
We propose a two dimensional frame-invariant phase field model of
grain impingement and coarsening.  One dimensional analytical
solutions for a stable grain boundary in a bicrystal are obtained, and
equilibrium energies are computed.  We are able to calculate the
rotation rate for a free grain between two grains of fixed
orientation.  For a particular choice of functional dependencies in
the model the grain boundary energy takes the same analytic form as the
microscopic (dislocation) model of Read and Shockley \cite{RS}.

\end{abstract}

\begin{multicols}{2}

    Previously we presented a new model \cite{KWC} and \cite{WCK}
    and phase-field simulations of the simultaneous processes of
    solidification, impingement, and coarsening of arbitrarily
    oriented crystals.  In earlier models of this phenomenon, a
    finite number of crystalline orientations are allowed with respect
    to the fixed coordinate reference frame. Morin et al.\cite{Morin},
    and Lusk \cite{Lusk1} constructed a free energy density having $N$
    minima by introducing a rotational (orientation) variable in the
    homogeneous free energy.  Chen and Yang\cite{Chen} and Steinbach
    et al.\cite{Steinbach} assigned $N$ order parameters to the $N$
    allowed orientations. In these approaches, the free energy density
    depends on the orientation of the crystal measured in the fixed
    frame---a property which is not physical.

Our previous work \cite{KWC} (KWC) introduced a model invariant under
rotations of the reference frame. A similarly motivated approach has
also been developed by Lusk \cite{Lusk2}. We have since determined
that the one dimensional solutions obtained in ~\cite{WCK} (WCK) are
globally unstable to a lower energy 
solution with completely diffuse grain boundaries.  Herein, we present
a completely general formulation of a new class of models which allow
for accurate, physical modeling of grain boundary formation by
impingement and subsequent motion.

In KWC, a rotationally invariant three-parameter phase field model of
solidification with subsequent impingement of grains and coarsening
was proposed.  The model parameters, $\phi$, $\eta$, and $\theta$,
represent: a liquid-solid order parameter, a coarse-grained measure of
the degree of crystalline order, and
the crystalline orientation, respectively.  For what follows we shall
focus on the events after impingement, when the material is completely
solid ($\phi=1$), since we are interested in the construction of a
physically realistic model of grain boundaries.  At the close of this
letter we will reintroduce solidification to complete the model.

We now more precisely define $\eta$ and $\theta$. Consider a fixed
subregion of solid material.  At the atomic scale we 
can define a discrete variable $\theta_{i}$  which
represents the orientation of an atomic bond (lattice vector) with
respect to some fixed laboratory frame. For this  subregion, we
define $\eta$ and $\theta$ such that
\begin{eqnarray}
(\eta \cos{\theta}, \eta \sin{\theta}) 
    = \frac{1}{N} \sum_{i=1}^{N}(\cos{\theta_{i}}, \sin{\theta_{i}}),
\label{eq:definition}
\end{eqnarray}
 where $N$ is the number of bonds in the subregion. Then, the variable
$\eta$ is an order parameter for the degree of crystalline
orientational order, where $\eta = 1$ signifies a completely oriented
state and $\eta = 0$ a state where no meaningful value of orientation
exists in the subregion, and $0 \le \eta \le 1$ always holds.  The
variable $\theta$ is an indicator of the mean orientation of the
crystalline subregion. Note, that when symmetrically equivalent
crystalline directions exist, $\theta$ may not be uniquely
defined. However, the point group symmetry operations can be used to
map each of the $\theta_i$ into a subdomain of $(0, 2\pi)$ where they
are unique.

Given the order parameters $\eta$ and $\theta$ we need to construct a
 free energy which is  invariant under rotations of the reference frame.
We start from the expression
\begin{equation} F=\int 
\left[f(\eta)+s \mu(\eta) G(|\nabla\theta|)+{\nu^2\over2}
|\nabla\eta|^2\right]dV,
\end{equation}
where $\nu$ and $s$ are constants, $f(\eta)$ is the homogeneous free
 energy and $G(|\nabla\theta|)$ is specified below. The 
 function $\mu(\eta)$ should have the property $\mu(0) = 0$
and  should increase monotonically with $\eta$.
The constants $s$ and $\nu$ are parameters which set the strength
of the penalties for gradients in misorientation and degree of crystalline
orientation, respectively.
The total free energy is frame invariant since it includes
the variable $\theta$ only in the form $\nabla\theta$.
For this work we assume that only ordered phase is stable, so $f(\eta)$ has a 
single-well at $\eta = 1$.

In two dimensions, the equilibrium grain boundary energy is generally
a function of two angles, for example, the angles of a boundary with
respect to the grains on either side.  We will derive the equilibrium
grain boundary energy as a function of the difference of these angles
(the misorientation), $\Delta \theta$, in one dimension.  Effects of
anisotropy (the dependence of grain boundary energy with respect to
boundary orientation; i.e., the direction of $\nabla \eta$ with
respect to mean crystalline orientation $\bar{\theta}$) are not
currently included in our model.

Because the energy form does not include $\theta$ explicitly,
no preferred values of $\theta$ are selected by the form of our
equations.
Therefore,
values of $\theta$ are determined by Dirichlet boundary conditions,
if they exist, and by gradient flow if the value of $\theta$ is
not fixed at any spatial position.
For a bicrystal (a single grain boundary) 
with Dirichlet boundary conditions, the spatial extent
of the grain boundary is determined by $G(|\nabla\theta|)$ and
the coupling to $\eta$.
The customary choice of square gradient,
$G(|\nabla\theta|) = \nabla \theta \cdot \nabla \theta$,
{\em cannot} be physically correct, since it leads to a
diffusion equation for $\theta$ and therefore the region of misorientation
spreads indefinitely.
In WCK an attempt
to stabilize a finite grain boundary was made by including many minima
in $G$
at different values of $|\nabla\theta|$.  With 
this form of $G$, a one dimensional solution to the equations of
motion can be found which has a finite grain boundary thickness.
We demonstrate below that this solution has a larger
energy than a completely diffuse interface, and is thus unstable.

It is useful to
 determine the energetic contribution  of terms with the form
 $|\nabla\theta|^{\alpha}$; where $\alpha$ is a positive number.  
 If the gradient is completely spread out over a one dimensional  domain 
 of length $L$ 
 and $\eta$ is uniform, then $\nabla\theta \equiv \Delta\theta / L$, 
 where $\Delta\theta$ is the change in
 angle across the domain (grain boundary misorientation). 
 If $\eta \equiv {\bar \eta}$ where ${\bar \eta}$ gives a 
 minimum of $f(\eta) + s \mu(\eta) |\nabla \theta|^{\alpha}$, 
and $L$ is taken to be large, the total excess energy (associated with
 $\Delta\theta\neq 0$) is given by 
\begin{eqnarray}
\Delta F_{\alpha, L} = \int_{-\frac{L}{2}}^{\frac{L}{2}} \left[
\Delta f({\bar \eta}) + s \mu({\bar \eta})|\nabla \theta|^{\alpha} 
\right] dx = O(L^{1-\alpha}),\nonumber
\end{eqnarray}
where $\Delta f( \eta)\equiv f(\eta)-f(1)$. In particular, for $\alpha
 > 1$ 
\begin{eqnarray}
\lim_{L\to\infty} \Delta F_{\alpha, L} = 0.
\end{eqnarray}
 Thus, there is {\sl no} form for $\alpha > 1$ 
 leading order behavior in $G$ which has a globally stable,
 non-diffuse solution, since a completely diffuse boundary has an
 energy which can always be made smaller than any other configuration.

However, the form of $\Delta F_{\alpha,L}$ suggests that the choice of 
$\alpha = 1$ may admit solutions for equilibrium grain boundaries of finite
spatial extent. 
If $\alpha < 1$ then the  derived evolution equation is completely singular
in the region where $\nabla\theta = 0$. Thus, $\alpha = 1$ 
($G(|\nabla \theta|) = |\nabla \theta|$ to leading order)
{\em is the only possible choice.}


Therefore, the dimensionless energy takes the following form:
\begin{eqnarray}
  F= \int_{\Omega} \left[
      f(\eta)+ \frac{\nu^{2}}{2} |\nabla \eta|^{2} 
      + s \mu(\eta) |\nabla \theta| 
      \right] dV,
\label{eq:energy}
\end{eqnarray} 
Here we choose for $f(\eta)$ and $\mu(\eta)$ the forms
\begin{equation}
  f(\eta)=\frac{1}{2} (1 - \eta)^{2};\ \ \   \mu(\eta) = \eta^2.
\end{equation}
We examine an alternative choice for $\mu(\eta)$ at the close 
of this letter.

We consider $\eta$ and $\theta$ as a polar coordinate system 
on the unit disk $D = \{\eta \le 1\}$ as Eqn.\ref{eq:definition} 
suggests, and introduce $L^{2}$-norm to the functional
space \cite{CarterCahnTaylor}.
Then, the equations of motion
are
\begin{eqnarray}
  \tau_{\eta} \eta_{t} 
  &=& \nu^{2} \nabla^2 \eta + 1 - \eta - 2 s \eta |\nabla \theta|,
\label{eq:evolve_eta}
\\ \nonumber
\\ 
  \tau_{\theta} \eta^{2} \theta_{t}
  &=& s \nabla \cdot
    \left[\eta^{2} \frac{\nabla \theta}{|\nabla \theta|}\right].
\label{eq:evolve_theta}
\end{eqnarray} where $\tau_{\eta}$ and $\tau_{\theta}$ are (possibly
anisotropic) inverse mobilities~\cite{KWC}. 

The form of Eqn.~\ref{eq:evolve_theta} requires careful
consideration since it contains a singular diffusivity
where $\nabla \theta=0$.  However, this singularity can be 
treated rigorously and controlled, as the work of
Giga and Giga \cite{GG} and Kobayashi and Giga \cite{KG} has indicated.  
A complete discussion of
the dynamics of this model is reserved for a longer paper, here we
present equilibrium one-dimensional solutions.
 
The equilibrium solutions $\eta(x)$ and $\theta(x)$ must satisfy the 
following equations:
\begin{eqnarray}
  0 &=& \nu^{2}\eta_{xx} + 1 - \eta - 2 s \eta |\theta_{x}|,
\label{eq:eta_1D_stat}
\\ \nonumber
\\
  0 &=& s \left[\eta^{2} \frac{\theta_{x}}{|\theta_{x}|}\right]_{x},
\label{eq:theta_1D_stat}
\end{eqnarray}
where the subscript $x$ indicates differentiation with respect
to $x$.
Dirichlet boundary conditions
$\theta (\pm\infty) =\theta_{\pm}$ are applied as well as the
condition $0 \leq \eta(x) \leq 1$.
Here, let us consider the equilibrium solution which corresponds 
to a bicrystal.
Without loss of generality, the center of the grain boundary is
located at $x = 0$.
If $\eta(x)$ has only one minimum, it can be  shown 
that the solution for $\theta(x)$ 
is a step function at the point where $\eta$ takes minimum
\cite{KG}. 
So we can take $\theta$ as
\begin{equation}
\theta(x) = \left\{
\begin{array}{lr}
\theta_{-} & \hspace{5mm} -\infty < x < 0;\\
\theta_{+} & \hspace{5mm} 0 < x < \infty.
\end{array}
\right.
\label{eq:step_theta}
\end{equation}
 Note that $|\theta_{x}| = \Delta\theta\;\delta(x)$, where 
$\Delta\theta = |\theta_{+} - \theta_{-}|$ and $\delta(x)$ 
is the Dirac delta function.
Eqn. \ref{eq:eta_1D_stat} gives
\begin{equation}
  \eta(x) = 1 - (1 - \eta_{0}) e^{-\frac{|x|}{\nu}}
\label{eq:stat_eta}
\end{equation} 
where $\eta_{0} = \eta(0)$, and should be determined by
integration of Eqn. \ref{eq:eta_1D_stat}
through the discontinuity in $\theta$:
\begin{equation}
  \nu^{2} \left[\eta_x\right]_{x=0^-}^{x=0^+} = 2 s \eta_0 \Delta\theta.
\label{eq:stat_eta_jump}
\end{equation}
Finally, Eqns. \ref{eq:stat_eta} and \ref{eq:stat_eta_jump} determine
$\eta_0$ as 
\begin{eqnarray}
  \eta_0 &=& {1\over 1+\Theta};\ \ \Theta \equiv {s\Delta\theta\over \nu},
\label{eq:eta_solution}
\end{eqnarray}
where $\Theta$ is a scaled misorientation.

We now calculate the total free energy.
This quantity can be separated into two terms:
the bulk contribution $E_{bulk}$, and the contribution from
the singular core at $x=0$,  $E_{core}$.
\begin{equation}
  E_{bulk} =
  \int_{-\infty}^{\infty}\left[
     \frac{\nu^{2}}{2}\eta_{x}^{2} 
   + \frac{1}{2} (1 - \eta)^{2} \right] dx
 = \nu \left({\Theta\over 1+\Theta}\right)^{2}
\label{eq:energy_normal}
\end{equation}
and
\begin{eqnarray}
E_{core} = s \eta_{0}^{2} \Delta\theta
      = \nu \Theta \left({1\over 1+\Theta}\right)^2,
\label{eq:energy_singular}
\end{eqnarray}
With these two energies we can find the surface excess energy
\begin{equation}
\gamma \equiv E_{bulk}+E_{core}={\nu\Theta\over 1+\Theta}.
\label{eq:energy_exact}
\end{equation}
Note that the energy is localized in the $\nu$-neighborhood of
the grain boundary,  and the precise contribution  to the energy from
the region $|x| > \ell$  is $E_{bulk}e^{-\frac{2\ell}{\nu}}$.

\begin{figure}[hbt]
\hbox to\hsize{\epsfxsize=1.0\hsize\hfil\epsfbox{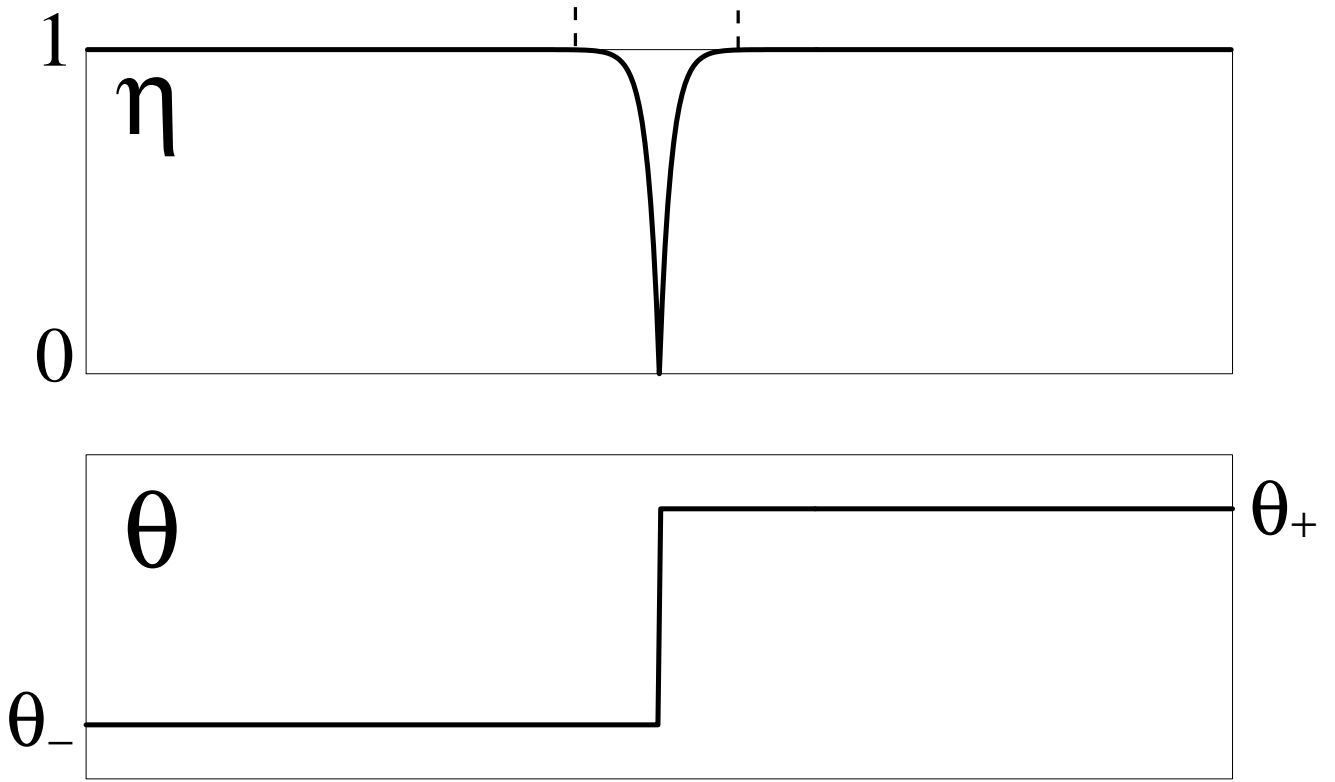}\hfil}
\nobreak\bigskip
\caption{Typical profiles of $\eta$ and $\theta$ with one grain boundary.
Simulated solution and analytic solution are both drawn, which coincide
completely within the resolution of graph. Domain size $= 1.0$, 
$\nu = 0.01$, $s = 100$ and $\Delta\theta = \frac{2\pi}{3}$.}
\label{fig:typical}
\end{figure}

When
\begin{eqnarray}
  s \gg \nu,
\label{eq:assump}
\end{eqnarray}
then, for almost all $\Delta\theta$, we have that $\Theta \gg 1$,
since $\Delta\theta$ is 
expected to have the values of order one. Therefore, except for 
small values of $\Delta\theta$, we have
\begin{equation}
  E_{bulk} \simeq \nu;\ \ \ E_{core} \simeq \nu \Theta^{-1} \ll E_{bulk}
\end{equation}
and
\begin{eqnarray}
  \gamma \simeq E_{bulk} \simeq \nu.
\label{eq:E_value}
\end{eqnarray}
Also, 
\begin{eqnarray}
  \eta_0 \simeq \Theta^{-1} \ll 1,
\end{eqnarray}
which means the degree of orientational order almost vanishes at the grain 
boundary.
The typical profiles of $\eta$ and $\theta$ are shown in
Fig \ref{fig:typical}.
Eqn. \ref{eq:E_value} implies that the grain boundary energy is
independent of misorientation provided that $\Delta \theta$ is 
much larger than the resolution level ${\nu}/{s}$ which is very small
in this particular case.
Therefore, we can regard the case $s \gg \nu$ to be independent of
misorientation.

\begin{figure}[hbt]
\hbox to\hsize{\epsfxsize=1.0\hsize\hfil\epsfbox{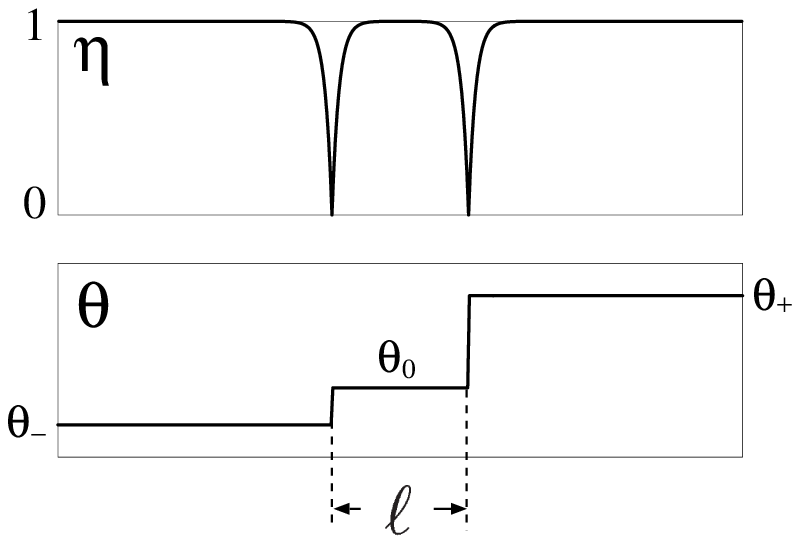}\hfil}
\nobreak\bigskip
\caption{A schematic plot of the order parameter $\eta$ and
orientation $\theta$ 
for a small grain with orientation $\theta_0$ trapped between two
pinned grains with orientation $\theta_\pm$. }
\label{fig:multi}
\end{figure}

So far, we have concentrated on the one grain boundary solution
under Dirichlet boundary conditions.
If the initial conditions had more than one boundary,
the unconstrained grain(s) must rotate to lower the
energy by adjusting the two boundary misorientations,
until the equilibrium solution obtained above is attained.
Since we are ultimately interested in the
dynamics of this problem, it is useful to examine the case
shown in Fig.\ref{fig:multi}, where a narrow band-shaped
interior grain with thickness $\ell$ lies between two semi-infinite grains. 
Like before, Dirichlet conditions are imposed as $\theta(\pm\infty) =
\theta_{\pm}$.  

Let the orientation of the interior grain be labeled $\theta_{0}$, and
assume $\ell\gg\nu$, then the system is described by $\eta$ and  
$\theta$ as shown in Fig.\ref{fig:multi}.
The system can eliminate the interior grain by two independent mechanisms:
1) by moving the boundaries towards each other at fixed
misorientations;
2) by adjusting the orientation(s) of the interior grain(s) at
fixed boundary spacing.
Since the regions where $\eta(x)$ varies rapidly are spatially 
localized ($\ell \gg \nu$), variation of the boundary position has 
negligible effect.
Therefore, we evaluate the rotation rate of the interior grain.
If we assume, for example, $\theta_{-} < \theta_{0} < \theta_{+}$, 
the following estimate holds~\cite{KG}:
\begin{eqnarray}
\tau_{\theta}\frac{\partial \theta_0}{\partial t} \simeq
\frac{\nu}{\ell}\frac{\nu}{s}\frac{2(\theta_{+}-\theta_{-})}
{(\theta_{0}-\theta_{-})^{2}(\theta_{+}-\theta_{0})^{2}}
\left(\theta_{0} - \frac{\theta_{+}+\theta_{-}}{2} \right)
\label{eq:sensitivity_exact}
\end{eqnarray}
It follows from (\ref{eq:sensitivity_exact}) that 
the angle $\theta_{0}$ rotates to that $\theta_{\pm}$ 
which is closer to $\theta_{0}$.
If $\tau_\theta$ is identically 0 then the rotation
rate is infinite, and a multi-grained solution would disappear
instantly. 

As a final point, for much of this analysis we have concerned
ourselves with the case where $\Theta \gg 1$. If $\Delta\theta$ is so
small that $\Theta \ll 1$ holds, Eqn.\ref{eq:energy_exact} gives the
asymptotics
\begin{equation} \gamma \simeq s \Delta\theta. \label{gamma1} \end{equation}
The energies of  low-misorientation tilt grain boundaries have been
approximated  by summing the energy of distribution of edge
dislocations~\cite{RS}. Here we discuss a method for reproducing the
same energetic dependence on misorientation.  

It is straightforward to
repeat the one dimensional analysis given above for arbitrary
$f(\eta)$ and $\mu(\eta)$.  If this is done for the choice
 \begin{equation}
\mu(\eta)= -2\ln{(1-\eta)} - 2\eta,
\label{eq:q_rs}
\end{equation} 
in Eqn.~\ref{eq:energy} and leaving $f(\eta)$ unchanged, we find
$\eta$ is also given by Eqn.\ref{eq:stat_eta} 
and $\eta_{0}$ is determined by 
\begin{equation}
  (1 - \eta_{0})^{2} = \Theta \eta_{0}.
\label{eq:eta0_rs}
\end{equation}
This guarantees $0 \le \eta_{0} \le 1$ for arbitrary $\Theta \ge 0$,
and gives two asymptotic expansions of $\eta_{0}$:
\begin{eqnarray}
   \eta_{0} &=& \Theta^{-1} - 2\Theta^{-2} + \ldots
   \hspace{10mm}for\;\; \Theta \gg 1,\\
   \eta_{0} &=& 1 - \Theta^{\frac{1}{2}} + \ldots
   \hspace{18mm}for\;\; \Theta \ll 1.
\label{eq:asymptotics_rs}
\end{eqnarray}
Thus, for $\Theta\gg1$ we obtain the same expression as given by
Eqn. \ref{eq:E_value}. In  addition, for the very small values of
$\Delta\theta$  ($\Theta \ll 1$), 
\begin{equation}
  \gamma \simeq - s \Delta\theta \ln{\Delta\theta}
\label{eq:rs_gamma}
\end{equation}
holds.
Eqn. \ref{eq:rs_gamma} is the Read-Shockley energy of a low angle tilt
grain boundary \cite{RS}.   
This is a very useful result, as it allows us to mimic
the physical picture of a grain boundary as a collection of
dislocations using a macroscopic (coarse grained) free energy for
small misorientations.

This polycrystalline model can be adopted into traditional
elements of phase field modeling of
solidification; the full free energy will now be of the form
\begin{eqnarray} F=\int 
[&f(\phi,\eta,T,{c_i})+\Gamma^2(\eta,\nabla\phi,\theta/{\cal
S})\\ \nonumber
&+s \mu(\eta)|\nabla\theta|
 +{\nu^2\over2}|\nabla\eta|^2]dV.
\label{FreeFold}
\end{eqnarray}
The field $\phi$ is the liquid-solid order parameter, and the first
two terms include the functional dependencies necessary to describe
solidification with anisotropic kinetics and surface energy (see
\cite{Kob1}).  We include the temperature $T$, concentrations
$c_i$, and any other thermodynamic variable which would control the
state of the system. For anisotropic grain boundaries we need only
reformulate $\nu^2|\nabla\eta|^2$ to be functionally similar to
$\Gamma$.  With this complete model, the full process of grain
formation, collision and coarsening can be simulated.

To summarize, we have derived a class of models which can describe the
solidification, impingement and coarsening of grains. These models
are rotationally invariant, and can be modified in a straightforward
manner to include a variety of physics. All of these models are
analytically tractable in one dimension, which makes study of their
behavior particularly direct.   If we choose the coupling carefully we
are able to derive a grain boundary energy identical to the
Read-Shockley low misorientation tilt grain boundary energy. Following papers
will explore the dynamics of this model in one and two dimensions, and
include the effects of solidification, and grain boundary anisotropy.

Acknowledgments:
The authors are grateful for helpful discussions with
W.~J. Boettinger, J.~W. Cahn, F.~W. Gayle, Y. Giga, G.~B. McFadden, and
M.~T. Lusk.  RK was partially supported by the Grant-in-Aid of
Ministry of Education, Science and Culture of Japan, and partially
supported by NIST.


\end{multicols}

\begin{thebibliography}{99}

\bibitem{KWC}
R. Kobayashi, J. A. Warren, and W. C. Carter,  Physica D, {\bf 119},
415 (1998).

\bibitem{WCK}
J. A. Warren, W. C. Carter, and  R. Kobayashi accepted Physica A. 

\bibitem{Morin}
B. Morin, K. R. Elder, M. Sutton, and M. Grant,
Phys. Rev. Lett. {\bf 75}, 2156 (1995).  

\bibitem{Lusk1}
M. T. Lusk, Submitted to Physical Review B (1998)

\bibitem{Chen}
L. Q. Chen and W. Yang, Phys. Rev. B {\bf 50}, 15752 (1994).

\bibitem{Steinbach}
I. Steinbach, F. Pezzolla, B. Nestler, M. BeeBelber, R. Prieler, 
G.J.Schmitz, and J.L.L. Rezende,
Physica D {\bf 94}, 135 (1996).

\bibitem{Eyre}
David Eyre.
\newblock 1997.
\newblock private communication.

\bibitem{Lusk2}
M. T. Lusk, accepted Proc. Royal Soc. London. (1998)

\bibitem{GG}
M.-H. Giga, Y. Giga, Arch. Rational Mech. Anal., {\bf 141}, 117 (1998)

\bibitem{KG}
R. Kobayashi, Y. Giga, submitted to J. Stat. Phys.

\bibitem{RS} 
W. T. Read, and W. Shockley, Phys. Rev, {\bf 78}, 275 (1950)

\bibitem{Kob1}
R. Kobayashi, Physica D {\bf 63}, 410 (1993).

\bibitem{CarterCahnTaylor}
W. C. Carter, J. E. Taylor, and J. W. Cahn J. of Metals, {\bf 49}, 30 (1997)
\end{thebibliography}
\end{document}